\RequirePackage{fixltx2e}
\RequirePackage{fix-cm}

\documentclass[%
floatfix,
showkeys,
twocolumn, %
nofootinbib, %
superscriptaddress, %
longbibliography, %
balancelastpage, %
flushbottom, %
]{revtex4-1}

\usepackage{cmap}

\usepackage{ucs}
\usepackage[utf8x]{inputenc}
\usepackage[T1,T2A]{fontenc}
\usepackage[german,russian,english]{babel}

\usepackage[sort&compress]{natbib}
\usepackage{amsmath}
\usepackage{amssymb}

\usepackage{mathtools}
\mathtoolsset{
showonlyrefs,
mathic = true
}

\allowdisplaybreaks

\usepackage{hyperref}
\hypersetup{backref,
 colorlinks=false}
\hypersetup{pdfborder=0 0 0}

\usepackage{breakurl}

\usepackage{microtype}
\UseMicrotypeSet[protrusion]{alltext}

\usepackage{fixltx2e}

\usepackage{nicefrac}

\makeatletter
\def\ps@pprintTitle{%
     \let\@oddhead\@empty
     \let\@evenhead\@empty
     \let\@oddfoot\@empty
     \let\@evenfoot\@oddfoot}
\makeatother

\usepackage{comment}

\usepackage{cool}
\usepackage{physics}
\usepackage{tensor}

\renewcommand{\i}{\mathrm{i}}

\newcommand{\setN}{\mathbb{N}}
\newcommand{\setR}{\mathbb{R}}

\newcommand{\crd}[1]{\underline{\vphantom{j}{#1}}}
\begin{document}
\graphicspath{{image/sdu-doi-general/en/}}

\title{Operator Approach to the Master Equation for the One-Step Process}

\author{M. Hnatich}
\email{hnatic@saske.sk}
\affiliation{Bogoliubov Laboratory of Theoretical Physics\\
Joint Institute for Nuclear Research\\
Joliot-Curie 6, Dubna, Moscow region, 141980, Russia}

\author{E. G. Eferina}
\email{eg.eferina@gmail.com}
\affiliation{Department of Applied Probability and Informatics\\
Peoples' Friendship University of Russia\\
Miklukho-Maklaya str. 6, Moscow, 117198, Russia}

\author{A. V. Korolkova}
\email{akorolkova@sci.pfu.edu.ru} 
\affiliation{Department of Applied Probability and Informatics\\
Peoples' Friendship University of Russia\\
Miklukho-Maklaya str. 6, Moscow, 117198, Russia}

\author{D. S. Kulyabov}
\email{yamadharma@gmail.com}
\affiliation{Department of Applied Probability and Informatics\\
Peoples' Friendship University of Russia\\
Miklukho-Maklaya str. 6, Moscow, 117198, Russia}
\affiliation{Laboratory of Information Technologies\\
Joint Institute for Nuclear Research\\
Joliot-Curie 6, Dubna, Moscow region, 141980, Russia}

\author{L. A. Sevastyanov}
\email{leonid.sevast@gmail.com}
\affiliation{Department of Applied Probability and Informatics\\
Peoples' Friendship University of Russia\\
Miklukho-Maklaya str. 6, Moscow, 117198, Russia}
\affiliation{Bogoliubov Laboratory of Theoretical Physics\\
Joint Institute for Nuclear Research\\
Joliot-Curie 6, Dubna, Moscow region, 141980, Russia}

\begin{abstract}

\begin{description}
\item[Background]
Presentation of the probability as an intrinsic property of the nature
leads researchers to switch from deterministic to stochastic
description of the phenomena.  On the basis of the ideology of
N.\,G.~van~Kampen and C.\,W.~Gardiner, the procedure of stochastization
of one-step process was formulated.  It allows to write down the
master equation based on the type of of the kinetic equations
(equations of interactions) and assumptions about the nature of the
process (which may not necessarily by a birth--death process).  The
kinetics of the interaction has recently attracted attention because
it often occurs in the physical, chemical, technical, biological,
environmental, economic, and sociological systems.  However, there are
no general methods for the direct study of this equation.  The
expansion of the equation in a formal Taylor series (the so called
Kramers--Moyal's expansion) is used in the procedure of
stochastization of one-step processes.  It is also possible to apply
system size expansion (van Kampen's expansion).  Leaving in the
expansion terms up to  the second order we can get the Fokker--Planck
equation, and thus the Langevin equation.  It should be clearly
understood that these equations are approximate recording of the
master equation.

\item[Purpose]
However, this does not eliminate the need for the study of the master
equation. Moreover, the power series produced during the master equation
decomposition may be divergent (for example, in
spatial models). This makes it impossible to apply the classical
perturbation theory.

\item[Method]
It is proposed to use quantum field perturbation theory for the
statistical systems (the so-called Doi method).  The perturbation series
are treated in the spirit of the Feynman path integral, where the Green's
functions of the perturbed Liouville operator of the master equation
are propagators.  For more convenience of selection of the perturbed
and unperturbed parts of the Liouville operator and to obtain the
explicit form of the Green function of the master equation we need to
rewrite the equation in the occupation number representation (Fock
state).

\item[Results]
This work is a methodological material that describes the principles
of master equation solution based on quantum field
perturbation theory methods.  The characteristic property of the 
work is that it is intelligible for non-specialists in quantum
field theory.  As an example the Verhulst model is used
because of its 
simplicity and clarity (the first order equation is independent of
the spatial variables, however, contains
non-linearity).

\item[Conclusions]
We show the full equivalence of the operator and
combinatorial methods of obtaining and study 
of the one-step process master equation.
\end{description}

\end{abstract}


\maketitle

\section{Introduction}

In order to construct stochastic models of the one-step 
processes~\cite{ef-kor-gev-kul-sev:vestnik-miph:2014-3}
(birth--death processes) the combinatorial methodology based on 
N.\,G.~van~Kampen~\cite{van-kampen:stochastic::en} and
C.\,W.~Gardiner~\cite{gardiner:stochastic::en} ideology was worked out. 
Under this methodology the master equation (for one-step processes) is 
derived by using the interaction schemes. The obtained master equation is further converted to the 
Fokker--Planck equation by expansion in formal series 
(Kramers--Moyal's decomposition)~\cite{gardiner:stochastic::en}. However, 
it is necessary to study the possibility of using this expansion for each type of process.

Thus, it is necessary not only to study the master equation but also to 
justify its expansion. It seems that the quantum perturbation theory best fits all the 
requirements.
  
There are two types of formalism which are generally used in the quantum perturbation theory: 
the path integral formalism and the formalism of second quantization (canonical formalism).
 It is worthy of note that it is a matter of taste which type of formalism to use. In 
 a number of works~\cite{Doi:1976:second_quantization, Doi:1976:stochastic_theory,
  Hnatich:2000:velocity-fluctuation, hnatich:2011:random_sources::en,
  hnatich:2011:anomalous_kinetics::en,
  Hnatich:2013:field-theoretic_technique} the possibility of using the formalism of the second 
  quantization for statistical tasks was studied. However, these articles are intended for      
  theoretical physicist that strongly limits the audience that could use the scientific results of the articles.    

The structure of the article is as follows. In the section \ref{sec:notation} 
basic notations and conventions are introduced. 
Section~\ref{sec:one-step} contains
a brief introduction to the method of randomization of one-step processes. 
Section~\ref{sec:fock} describes the algorithm of the one-step processes recording in 
terms of the occupation number representation. The master equation in the form of the 
Liouville operator equation is also presented.
In the section~\ref{sec:model} case study model for both the combinatorial and operator 
approaches is described. The equivalence of the combinatorial and the operator 
approaches is proved.

\section{Notations and conventions}
\label{sec:notation}

\begin{enumerate}

\item The abstract indices notation~\cite{penrose-rindler:spinors::en}
  is used in this work.  Under this notation a tensor as a whole
  object is denoted just as an index (e.g., $x^{i}$), components are
  denoted by underlined index (e.g., $x^{\crd{i}}$).

 \item We will adhere to the following agreements. Latin indices from
   the middle of the alphabet ($i$, $j$, $k$) will be applied to the space
   of the system state vectors. Latin indices from the beginning of
   the alphabet ($a$) will be related to the Wiener process space. Greek
   indices ($\alpha$) will set a number of different interactions in
   kinetic equations.

\end{enumerate}

\section{One--step processes stochastization}
\label{sec:one-step}

The one--step processes (also known as the birth--death processes) are Markov 
processes with continuous time, integer state of states the transition matrix of which 
allows only transitions between neighbouring states.

\subsection{Interaction schemes}

The system state is defined by the vector
$\varphi^{i} \in \setR^n$, where $n$~--- is system order
\footnote{For brevity, we denote the module over the field $\setR$
  just as  $\setR$.}. The operator
$I^{i}_{j} \in \setN^{n}_{0} \times \setN^{n}_{0}$ describes the state 
of the system before the interaction, the operator
$F^{i}_{j} \in \setN^{n}_{0} \times \setN^{n}_{0}$~---
after the interaction\footnote{The component dimension indices 
take values $\crd{i},\crd{j} = \overline{1,n}$.}. The result of the interaction 
is the system transition from one state to another one.

There are $s$ types of interaction in our system, so instead of 
 $I^{i}_{j}$ and $F^{i}_{j}$ operators we will use operators
$I^{i \alpha}_{j} \in \setN^{n}_{0} \times \setN^{n}_{0} \times
\setN^{s}_{+}$
and
$F^{i \alpha}_{j} \in \setN^{n}_{0} \times \setN^{n}_{0} \times
\setN^{s}_{+}$\footnote{The component indices of number of interactions 
take on values $\crd{\alpha} = \overline{1,s}$}.

The interaction of the system elements will be described by interaction
schemes which are similar to the
schemes of the chemical kinetics:
\begin{equation}
  \label{eq:chemkin}
  I^{i \alpha}_{j} \varphi^j
  \overset{\tensor*[^{+}]{k}{_{\alpha}}}{\underset{\tensor*[^{-}]{k}{_{\alpha}}}{\rightleftharpoons}}
  F^{i \alpha}_{j} \varphi^{j},
\end{equation}
the Greek indices specify the number of interactions and the Latin ones 
the system order. The coefficients $\tensor*[^{+}]{k}{_{\alpha}}$ and
$\tensor*[^{-}]{k}{_{\alpha}}$ have the meaning of intensity (speed) of the interaction.

The state transition is given by the operator:
\begin{equation}
  \label{eq:r_i}
  r_j^{i \alpha} = F_j^{i \alpha} -I_j^{i \alpha}.
\end{equation}

Thus, the one step interaction $\crd{\alpha}$ in the forward and reverse directions can be written as
\begin{equation}
  \begin{gathered}
    \varphi^{i}  \rightarrow \varphi^i + r^{i \crd{\alpha}}_{j} \varphi^{j},\\
    \varphi^{i} \rightarrow \varphi^{i} - r^{i \crd{\alpha}}_{j} \varphi^{j}.
  \end{gathered}
\end{equation}

We can also write~\eqref{eq:chemkin} not as vector equations but as sums:
\begin{equation}
  \label{eq:chemkin2}
  I^{i \alpha}_{j} \varphi^j \delta_i
  \overset{\tensor*[^{+}]{k}{_{\alpha}}}{\underset{\tensor*[^{-}]{k}{_{\alpha}}}{\rightleftharpoons}}
  F^{i \alpha}_{j} \varphi^{j} \delta_i,
\end{equation}
where $\delta_{\crd{i}} = (1,\ldots,1)$.

Also the following notation will be used:
\begin{equation}
  \label{eq:n^i-notion}
  I^{i \alpha} := I^{i \alpha}_{j} \delta^{j}, 
  \quad F^{i \alpha} := F^{i \alpha}_{j} \delta^{j}, \quad
  r^{i \alpha} := r^{i \alpha}_{j} \delta^{j}.
\end{equation}

\subsection{The master equation}

For the system description we will use the master equation, which 
describes the transition probability for a Markov
process~\cite{van-kampen:stochastic::en, gardiner:stochastic::en}:
\begin{multline}
  \label{eq:master:trans}
  \pdv{p(\varphi_{2},t_{2}|\varphi_{1},t_{1})}{t} = \int \biggl[
  w(\varphi_{2}|\psi,t_{2}) p(\psi,t_{2}|\varphi_{1},t_{1}) 
  - {} \\ {} -
  w(\psi|\varphi_{2},t_{2}) p(\varphi_{2},t_{2}|\varphi_{1},t_{1}) 
  \biggr] \dd{\psi},
\end{multline}
where $w(\varphi|\psi,t)$ is the probability of transition from the state $\psi$
to the state $\varphi$ for unit time.

Fixing the initial values of $\varphi_{1},t_{1}$, we can write the equation for subensemble:
\begin{equation}
  \label{eq:master:subansemble}
  \pdv{p(\varphi,t)}{t} = \int
  \qty[
  w(\varphi|\psi,t) p(\psi,t) -
  w(\psi|\varphi,t) p(\varphi,t)
  ] \dd{\psi}.
\end{equation}

If the domain of variation of $\varphi$ is discrete, then the master equation~\eqref{eq:master:subansemble}
can be written as follows
 (the states are numbered by $n$ and $m$):
\begin{equation} 
  \label{eq:mas_eq}
    \pdv{p_{n}(t)}{t} = \sum\limits_{m} 
    \qty[w_{nm} p_{m}(t) - w_{mn} p_{n}(t)],
\end{equation}
where $p_{n}$ is the probability to find the system in the state $n$ at the time $t$,
$w_{nm}$ is the probability of the transition from the state $m$ to the state $n$ per unit time.

There are two types of system transitions from one state to another (based on one--step processes) 
as a result of system elements interaction: in the forward direction 
($\varphi^{i} + r^{i \crd{\alpha}}_{j} \varphi^{j}$) with probability
$\tensor*[^{+}]{s}{_{\crd{\alpha}}}(\varphi^{k})$ and in the opposite direction
($\varphi^{i} - r^{i \crd{\alpha}}_{j} \varphi^{j}$) with probability
$\tensor*[^{-}]{s}{_{\crd{\alpha}}}(\varphi^{k})$~(fig.~\ref{fig:one-step_process}). 
The matrix of the transition probabilities has the form: 
\begin{equation}
  w_{\crd{\alpha}}(\varphi^{i}| \psi^{i} ,t) = \tensor*[^{+}]{s}{_{\crd{\alpha}}}
  \delta_{\varphi^{i},\psi^{i}+1} + \tensor*[^{-}]{s}{_{\crd{\alpha}}} \delta_{\varphi^{i}, \psi^{i}-1},
\end{equation}
where $\delta_{i,j}$ is Kronecker delta.

Thus, the general form of the master equation for the state vector $\varphi^{i}$, 
changing by steps of length $r^{i \crd{\alpha}}_j \varphi^j$, is:
\begin{multline} 
  \label{eq:one-step:mu}
  \pdv{p(\varphi^{i} ,t)}{t} =  \sum_{\crd{\alpha}=1}^{s} 
  \Biggl\{
  \biggl[ \tensor*[^{-}]{s}{_{\crd{\alpha}}} (\varphi^{i}+r^{i
    \crd{\alpha}},t) p(\varphi^{i}+r^{i \crd{\alpha}} ,t) 
  - {} \\ - {}
  \tensor*[^{+}]{s}{_{\crd{\alpha}}}(\varphi^{i}) p(\varphi^{i},t)
  \biggr] 
  + {} \\ + {}
  \biggl[ \tensor*[^{+}]{s}{_{\crd{\alpha}}} (\varphi^{i}-r^{i
    \crd{\alpha}}, t) 
  p(\varphi^{i} - r^{i \crd{\alpha}},t) -
  \tensor*[^{+}]{s}{_{\crd{\alpha}}} (\varphi^i) p(\varphi^{i},t)
  \biggr] \Biggr\}.
\end{multline}

\begin{figure*}
  \centering
  \includegraphics[width=0.8\textwidth]{markov}
  \caption{One--step process}
  \label{fig:one-step_process}
\end{figure*}

The transition rates $\tensor*[^{+}]{s}{_{\crd{\alpha}}}$ and $\tensor*[^{-}]{s}{_{\crd{\alpha}}}$ 
are proportional to the number of ways of choosing the number of
arrangements of $\varphi^{\crd{i}}$ to $I^{\crd{i} \crd{\alpha}}$
(denoted as $A_{\varphi^{\crd{i}}}^{I^{\crd{i} \crd{\alpha}}}$) and to
$F^{\crd{i} \crd{\alpha}}$ (denoted as $A_{\varphi^{\crd{i}}}^{F^{\crd{i} \crd{\alpha}}}$) and defined by:
\begin{equation}
  \label{eq:s-pm}
\begin{gathered}
  \tensor*[^{+}]{s}{_{\crd{\alpha}}} =
  \tensor*[^{+}]{k}{_{\crd{\alpha}}} \prod_{\crd{i}=1}^{n} 
  A_{\varphi^{\crd{i}}}^{I^{\crd{i} \crd{\alpha}}} =
  \tensor*[^{+}]{k}{_{\crd{\alpha}}} \prod_{\crd{i}=1}^{n}
  \frac{\varphi^{\crd{i}}!}{(\varphi^{\crd{i}} - I^{\crd{i} \crd{\alpha}})!}, \\
  \tensor*[^{-}]{s}{_{\crd{\alpha}}} = 
  \tensor*[^{-}]{k}{_{\crd{\alpha}}} \prod_{\crd{i}=1}^{n} 
  A_{\varphi^{\crd{i}}}^{F^{\crd{i} \crd{\alpha}}} =
  \tensor*[^{-}]{k}{_{\crd{\alpha}}} \prod_{\crd{i}=1}^{n}
  \frac{\varphi^{\crd{i}}!}{(\varphi^{\crd{i}}-F^{\crd{i} \crd{\alpha}})!}.
\end{gathered}
\end{equation}

Replacing in~\eqref{eq:s-pm} the 
$\varphi (\varphi-1) \cdots (\varphi - (n-1))$-type combinations  on $(\varphi)^n$
we obtain for Fokker--Planck equation\footnote{This change corresponds to a series expansion.}:
\begin{equation}
  \label{eq:s-pm:exp}
\begin{gathered}
  \tensor*[^{+}_{\text{fp}}]{s}{_{\crd{\alpha}}} =
  \tensor*[^{+}]{k}{_{\crd{\alpha}}} \prod_{\crd{i}=1}^{n}
  (\varphi^{\crd{i}})^{I^{\crd{i} \crd{\alpha}}}, \\
  \tensor*[^{-}_{\text{fp}}]{s}{_{\crd{\alpha}}} = 
  \tensor*[^{-}]{k}{_{\crd{\alpha}}} \prod_{\crd{i}=1}^{n}
  (\varphi^{\crd{i}})^{F^{\crd{i} \crd{\alpha}}}.
\end{gathered}
\end{equation}

\subsection{Fokker--Planck equation}

Fokker - Planck equation is a special case of the master equation and can be regarded as its approximation.

We will use the decomposition of the Kramers--Moyal~\cite{gardiner:stochastic::en} (for simplicity it is 
written for the one-dimensional case):
\begin{equation}
  \label{eq:kramers-moyal}
  \pdv{p(\varphi,t)}{t} = \sum \limits_{n=1}^\infty \frac{(-1)^n}{n!}
  \pdv[n]{}{\varphi} \Big[\xi^n(\varphi) p(\varphi,t)\Big],
\end{equation}
where
\begin{equation}
  \xi^n(\varphi) = \int\limits_{-\infty}^{\infty} (\psi - \varphi)^{n}
  w(\psi|\varphi) \dd{\psi}.
\end{equation}

By dropping the terms with order higher than the second one, we obtain the Fokker--Planck equation:
\begin{equation}
  \label{eq:fp_1D}
  \pdv{p(\varphi, t)}{t} = 
  - \pdv{\varphi} \left[ A(\varphi) p(\varphi, t) \right] +
  \pdv[2]{}{\varphi} 
  \left[ B(\varphi) p(\varphi, t) \right],
\end{equation}
and for multivariate case
\begin{multline}
  \label{eq:fp_nD}
  \pdv{p(\varphi^k, t)}{t} = 
  - \pdv{}{\varphi^{i}} \left[ A^{i}(\varphi^{k}) p(\varphi^{k}, t)
  \right] 
  + {} \\ {} +
  \frac{1}{2} \pdv{}{\varphi^{i}}{\varphi^{j}} 
  \left[ B^{i j} (\varphi^k) p(\varphi^{k}, t)  \right],
\end{multline}
where
\begin{equation} 
  \label{eq:fp_coeff}
  \begin{gathered}
    A^{i} := A^{i}(\varphi^{k}) = r^{i \crd{\alpha}} \left[
      \tensor*[^+_{\text{fp}}]{s}{_{\crd{\alpha}}} -
      \tensor*[^-_{\text{fp}}]{s}{_{\crd{\alpha}}} \right], \\
    B^{i j} := B^{i j}(\varphi^{k}) = r^{i \crd{\alpha}} r^{j
      \crd{\alpha}} \left[ \tensor*[^+_{\text{fp}}]{s}{_{\crd{\alpha}}} -
      \tensor*[^-_{\text{fp}}]{s}{_{\crd{\alpha}}} \right].
  \end{gathered}
\end{equation}

As can be seen from the~\eqref{eq:fp_coeff}, the coefficients of the 
Fokker--Planck equation can be obtained directly from the~\eqref{eq:r_i} and
\eqref{eq:s-pm}, that is, in this case, it  is not necessary to write down the master equation. 

\subsection{Langevin equation}

The Langevin equation which corresponds to Fokker--Planck equation:
\begin{equation}
  \label{eq:langevin}
  \dd \varphi^{i} = a^{i} \dd{t} + b^i_{a} \dd{W^{a}},
\end{equation}
where $a^{i} := a^{i} (\varphi^k)$, $b^{i}_{a} := b^{i}_{a}
(\varphi^k)$, 
$\varphi^i \in \setR^n $ --- system state vector, $W^{a} \in \mathbb{R}^m$
--- $m$-dimensional Wiener process\footnote{Wiener process is realized as $\dd{W} = \varepsilon \sqrt{\dd{t}}$, 
where $\varepsilon \sim N(0,1)$~--- the normal distribution with mean $0$ and variation
$1$.}. Latin indices from the middle of the alphabet will be applied to the system state vectors 
(the dimensionality of space is $n$), and Latin indices from the beginning of 
the alphabet denote the variables related to the Wiener process vector 
(he dimensionality of space is $m \leqslant n$).

The connection between the equations \eqref{eq:fp_nD} and \eqref{eq:langevin}
is expressed by the following relationships:
\begin{equation}
  \label{eq:k-langevin}
  A^{i} = a^{i}, \qquad B^{i j} = b^{i}_{a} b^{j a}.
\end{equation}

\section{Occupation numbers representation} 
\label{sec:fock}

The occupation number representation is the main language in the description of many-body physics. 
The main elements of the language are the wave functions of the system, providing information 
about how many particles are in each single-particle state. 
The creation and annihilation operators are used for system states change. 
The advantages of this formalism are:
\begin{itemize}
\item  it is possible to  consider systems with a variable number of particles 
        (non-stationary systems);
\item the system statistics  (Fermi--Dirac or Bose--Einstein) is automatically included in 
        the commutation rules for the creation and annihilation operators;
\item this is the second major formalism (along with the path integral) for the quantum 
        perturbation theory description.
\end{itemize}

The application of the formalism of the second quantization to non-quantum systems 
(statistical, deterministic systems) was also studied ~\cite{Doi:1976:second_quantization,
  Doi:1976:stochastic_theory, zeldovich:1978:kinetics::en,
  grassberger:1980:fock-space,
  peliti:1985:path_integral_approach}.

The Dirac notation is commonly used for  occupation numbers representation recording.

\subsection{Dirac notation}

This notation is proposed by P.\,A.\,M. Dirac~\cite{dirac:1939:braket}\footnote{The notation 
is based on the notation, proposed by G. Grassmann in 1862~\cite[p.~134]{cajori:1929:history}.}. 
The vector $\varphi^{i}$ is denoted $\ket{i}$,
the covariant vector (covector) $\varphi_{i}$ is denoted
$\bra{i}$. The conjunction operation is used for raising and lowering the
indices\footnote{In this case, we use Hermitian conjugation ${\bullet}^{\dagger}$. 
The sign of the complex conjugate ${\bullet}^{*}$ in this entry is superfluous.}:
\begin{equation}
  \label{eq:dirac:index_conj}
  \varphi^{*}_{i} := \varphi_{i} = (\varphi^{i})^{\dagger} \equiv \bra{i} = \ket{i}^{\dagger}.
\end{equation}

The scalar product is: 
\begin{equation}
  \label{eq:dirac:inner}
  \varphi_{i} \varphi^{i} \equiv \bra{i}\ket{i}.
\end{equation}

The tensor product is:
\begin{equation}
  \label{eq:dirac:outer}
  \varphi_{j} \varphi^{i} \equiv \ket{i}\bra{j}.
\end{equation}

Another form of the Dirac notation is also possible: 
\begin{equation}
  \label{eq:dirac:wavefunc}
  \ket{\varphi} := \varphi^{i}, \qquad 
  \bra{i}\ket{\varphi} := \varphi^{i} \delta^{\crd{i}}_{i} = \varphi^{\crd{i}}.
\end{equation}

\subsection{Creation and annihilation operators}

The transition to the space of occupation numbers is not a unitary transformation. 
However, the algorithm of the transition (specific to each task) can be constructed.

This procedure is illustrated here for the master
equation~\eqref{eq:mas_eq} for a system which does not depend on the
spatial variables and is
one-dimensional.

The probability that the system of interest consists of $n$ particles is 
\begin{equation}
  \label{eq:phi_n}
  \varphi_{n} := p_{n} (\varphi, t).
\end{equation}

The vector space $\mathcal{H}$ consists of states of $\varphi$.

The scalar product can be written:
\begin{equation}
  \label{eq:scalar_prod_ex}
  \bra{\varphi}\ket{\psi} = \sum_n n! p_{n}^{*} (\varphi) p^{n}(\psi)
  = \sum_n n! \varphi_{n}^{*} (\varphi) \varphi^{n}(\psi)
\end{equation}
and $\ket{n}$ is a basis vectors.

From $p_{n}(m) = \delta_{n}^{m}$ and~\eqref{eq:scalar_prod_ex} we obtain:
\begin{equation}
  \label{eq:<n|m>}
\bra{n}\ket{m} = n!\delta_{n}^{m}.
\end{equation}

The state vector:
\begin{equation}
  \label{eq:phi_vector}
  \ket{\varphi} = \sum_{n} p_{n} (\varphi) \ket{n} = \sum_{n}
  \varphi_{n} \ket{n} =: \varphi_{n} \ket{n}.
\end{equation}

In view of~\eqref{eq:<n|m>}, we get:
\begin{equation}
  \label{eq:phi_n-scalar}
  \varphi_{n} = \frac{1}{n!} \bra{n}\ket{\varphi}.
\end{equation}

Creation and annihilation operators are defined respectively as:
\begin{equation}
  \label{eq:creat+annihil}
  \begin{gathered}
    \pi \ket{n} = \ket{n+1}, \\
    a \ket{n} = n \ket{n-1}.
  \end{gathered}
\end{equation}
They satisfy the commutation rule\footnote{In fact, $a\pi\ket{n} -
  \pi a \ket{n} = (n+1)\ket{n} - n \ket{n} = \ket{n}$.}:
\begin{equation}
  \label{eq:commutator}
  [a,\pi] = 1.
\end{equation}

From~\eqref{eq:scalar_prod_ex} and~\eqref{eq:commutator} it follows that the system is described 
by  Bose--Einstein statistics.  

From~\eqref{eq:<n|m>} we get: 
\begin{equation}
  \label{eq:scalar-complex-conj}
  \mel{m}{a^{\dagger}}{n} = \mel{m}{\pi}{n},
\end{equation}
and therefore: 
\begin{equation}
  \label{eq:a+=pi}
  a^{\dagger} = \pi.
\end{equation}

\subsection{Liouville operator}

The Liouville equation:
\begin{equation}
  \label{eq:liuville}
  \pdv{}{t} \ket{\varphi(t)} =  L \ket{\varphi(t)}.
\end{equation}
Liouville operator $L$ satisfies the relation:
\begin{equation}
  \label{eq:liuville=0}
  \bra{0} L = 0.
\end{equation}

From~\eqref{eq:mas_eq}, \eqref{eq:phi_vector}, \eqref{eq:phi_n-scalar}
and~\eqref{eq:liuville} we obtain:
\begin{multline}
  \label{eq:master2liuville}
   \pdv{p_{n}}{t} = \frac{1}{n!} \bra{n}\pdv{}{t}\ket{\varphi} =
   \frac{1}{n!} \bra{n}L\ket{\varphi} 
   = {} \\ {} =
   \sum\limits_{m} 
    \qty[w_{nm} p_{m} - w_{mn} p_{n}].
\end{multline}
In this way, the system of master equations~\eqref{eq:mas_eq} has been
reduced to a single equation, the Liouville
equation~\eqref{eq:liuville}.The following Liouville operator
corresponds to the
scheme~\eqref{eq:chemkin}:
\begin{multline}
  \label{eq:liuville:chemkin}
  L = \sum_{\crd{\alpha},\crd{i}} \biggl[
  \tensor*[^{+}]{k}{_{\crd{\alpha}}}
  \qty(
  (\pi_{\crd{i}})^{F^{\crd{i}\crd{\alpha}}} -
  (\pi_{\crd{i}})^{I^{\crd{i}\crd{\alpha}}} 
  ) (a_{\crd{i}})^{I^{\crd{i}\crd{\alpha}}}
  + {} \\ {} +
  \tensor*[^{-}]{k}{_{\crd{\alpha}}}
  \qty(
  (\pi_{\crd{i}})^{I^{\crd{i}\crd{\alpha}}} -
  (\pi_{\crd{i}})^{F^{\crd{i}\crd{\alpha}}} 
  ) (a_{\crd{i}})^{F^{\crd{i}\crd{\alpha}}}
  \biggr].
\end{multline}

\section{Verhulst model}
\label{sec:model}

As a demonstration of the method, we consider the Verhulst 
model~\cite{verhulst:1838, Feller:1939:acta_biotheoretica,
  feller:1949:theory_stochastic_processes}, which describes the limited 
  growth\footnote {The attractiveness of this model is that it is one-dimensional and non-linear.}. 
Initially, this model was written down as the differential equation:
\begin{equation}
    \dv{\varphi}{t}= \lambda \varphi - \beta \varphi - \gamma \varphi^{2},
\end{equation}
where $\lambda$ denotes the breeding intensity factor, $\beta$~--- the
extinction intensity factor, $\gamma$~--- the factor of population
reduction rate (usually the rivalry of individuals is
considered)\footnote{The same notation as in the original
  model~\cite{verhulst:1838} is used.}.

The interaction scheme for the stochastic version of the model is:
\begin{equation}
  \label{eq:verhulst:chemkin}
    \begin{gathered}
        \varphi \overset{\lambda}{\underset{\gamma}{\rightleftharpoons}} 2\varphi ,\\
        \varphi \xrightarrow{\beta} 0.
    \end{gathered}
    \qquad
    \begin{gathered}
      I^{\crd{i}\crd{\alpha}}=
      \begin{pmatrix}
        1 & 1 
      \end{pmatrix}, \\
      F^{\crd{i}\crd{\alpha}}=
      \begin{pmatrix}
        2 & 0 
      \end{pmatrix}.
    \end{gathered}
    \qquad
    \begin{gathered}
      r^{\crd{i}\crd{\alpha}} =
      \begin{pmatrix}
        1 & -1
      \end{pmatrix}.
    \end{gathered}
  \end{equation}

The first relation means that an individual who eats one unit of meal is immediately reproduced, 
and in the opposite direction is the rivalry between individuals. 
The second relation describes the death of an individual.

\subsection{One--step processes stochastization method}

The correspondence between the terms entering the Fokker--Plank
equation~\eqref{eq:s-pm:exp} and the transition rates defined within
the Verhults model is as follows:
\begin{equation}
  \begin{gathered}
    \tensor*[^{+}]{s}{_{1}} = \lambda \varphi, \\
    \tensor*[^{-}]{s}{_{1}} = \gamma \varphi (\varphi - 1), \\
    \tensor*[^{+}]{s}{_{2}} = \beta \varphi.
  \end{gathered}
  \qquad
  \begin{gathered}
    \tensor*[^{+}_{\text{fp}}]{s}{_{1}} = \lambda \varphi, \\
    \tensor*[^{-}_{\text{fp}}]{s}{_{1}} = \gamma \varphi^{2}, \\
    \tensor*[^{+}_{\text{fp}}]{s}{_{2}} = \beta \varphi.
  \end{gathered}
\end{equation}

Then, based on~\eqref{eq:one-step:mu}, the form of the master equation is:
\begin{multline}
  \label{eq:master}
  \pdv{p (\varphi,t)}{t} =  
  - \qty[\lambda \varphi + \beta \varphi +
  \gamma \varphi (\varphi -1)] 
  p(\varphi,t) 
  + {} \\ {} +
  \qty[\beta (\varphi + 1)
  +  \gamma (\varphi + 1) \varphi] 
  p(\varphi + 1,t) + \lambda (\varphi - 1) p (\varphi - 1,t).
\end{multline}

For particular values of $\varphi$ (as in~\eqref{eq:mas_eq}):
\begin{multline}
  \label{eq:verhulst:master:n}
  \pdv{p_{n}(t)}{t} := \pdv{p (\varphi,t)}{t}\eval_{\varphi=n} 
  = {} \\ {} =
  - \qty[\lambda n + \beta n +
  \gamma n(n-1)] 
  p_{n}(t) 
  + {} \\ {} +
  \qty[\beta (n + 1) + \gamma (n + 1)n] 
  p_{n+1}(t) + 
  \lambda (n - 1) p_{n-1} (t).
\end{multline}

Through the use of~\eqref{eq:fp_1D} the Fokker--Planck equation is obtained:
\begin{equation}
  \pdv{p(\varphi,t)}{t}=
  -\pdv{}{\varphi} \left(A p(\varphi,t)\right)+\frac{1}{2}
  \pdv[2]{}{\varphi} \left(B p(\varphi,t)\right),
\end{equation}

The coefficients $A$ and $B$ are equal (refer to~\eqref{eq:fp_coeff}):

\begin{equation}
  \begin{gathered}
    A = \lambda \varphi -\beta\varphi - \gamma \varphi^2, \\
   B = \lambda \varphi + \beta\varphi - \gamma \varphi^2.
  \end{gathered}
\end{equation}

From the Fokker--Planck equation the Langevin form of equivalent system
 of differential equations is obtained:
\begin{equation}
  \dd{\varphi(t)} = (\lambda \varphi -\beta\varphi - \gamma
  \varphi^2)\dd{t} +
 \sqrt {(\lambda \varphi + \beta\varphi - \gamma \varphi^2)} \dd{W(t)}.
\end{equation}

\subsection{Occupation number representation}

From~\eqref{eq:verhulst:chemkin} and~\eqref{eq:liuville:chemkin} the 
Liouville operator is:
\begin{multline}
  \label{eq:verhulst:liuville}
  L = \lambda (\pi^2 -\pi) a + \gamma (\pi - \pi^2) a^2 + 
  \beta (1 - \pi) a 
  = {} \\ {} =
  \lambda \qty((a^{\dagger})^2 - a^{\dagger}) a + 
  \gamma \qty(a^{\dagger} - (a^{\dagger})^2) a^2 + 
  \beta \qty(1 - a^{\dagger}) a 
  = {} \\ {} =
  \lambda \qty(a^{\dagger} - 1)a^{\dagger} a +
  \beta \qty(1 - a^{\dagger}) a + 
  \gamma \qty(1 - a^{\dagger})a^{\dagger} a^2.
\end{multline}

The master equation by Liouville operator (from~\eqref{eq:master2liuville}) and by means of~\eqref{eq:<n|m>}, \eqref{eq:creat+annihil},
\eqref{eq:phi_n} and~\eqref{eq:phi_n-scalar}:
\begin{multline}
  \label{eq:verhulst:master2liuville}
  \pdv{p_{n}(t)}{t} = \frac{1}{n!} \bra{n}L\ket{\varphi} 
  =  {} \\ {} =
  \frac{1}{n!} \bra{n}
  - \qty[ 
  \lambda a^{\dagger} a +
  \beta a^{\dagger} a +
  \gamma a^{\dagger}a^{\dagger} a a
  ]
  + {} \\ {} +
  \qty[
  \beta a + 
  \gamma a^{\dagger} a a
  ]
  +
  \lambda a^{\dagger}a^{\dagger} a 
  \ket{\varphi}
  = {} \\ {} =
  - \qty[\lambda n + \beta n +
  \gamma n(n-1)] 
  \bra{n}\ket{\varphi}
  + {} \\ {} +
  \qty[\beta (n + 1) + \gamma (n + 1)n] 
  \bra{n+1}\ket{\varphi} + 
  \lambda (n - 1) \bra{n-1}\ket{\varphi}
  = {} \\ {} =
  - \qty[\lambda n + \beta n +
  \gamma n(n-1)] 
  p_{n}(t) 
  + {} \\ {} +
  \qty[\beta (n + 1) + \gamma (n + 1)n] 
  p_{n+1}(t) + 
  \lambda (n - 1) p_{n-1} (t).
\end{multline}

The result~\eqref{eq:verhulst:master2liuville} coincides with 
the formula~\eqref{eq:master}, which was obtained by combinatorial method.

\section{Conclusions}
\label{sec:conclusion}

This article introduced the operator method for one--step processes. 
At all stages of the operator method it is compared with the 
combinatorial method of stochastization of the one--step processes. The
logic of both methods 
is demonstrated. The complete equivalence of both methods is presented by their comparison.

However, at this stage it is a difficult task to justify a preference 
for one of the methods.  But it should be noted that the operator
formalism allows to use the achievements made within the framework of the
quantum field theory in a more familiar way.

\begin{acknowledgments}

The work is partially supported by RFBR grants No's 14-01-00628 and
15-07-08795.

Published in:
M.~Hnati{\v{c}}, E.~G. Eferina, A.~V. Korolkova, D.~S. Kulyabov, L.~A.
Sevastyanov,
Operator
Approach to the Master Equation for the One-Step Process, EPJ Web of
Conferences 108 (2016) 02027.
\href{http://dx.doi.org/10.1051/epjconf/201610802027}{doi:10.1051/epjconf/201610802027}.

Sources:
\url{https://bitbucket.org/yamadharma/articles-2014-sdu-doi}

\end{acknowledgments}

  \bibliographystyle{elsarticle-num}

\bibliography{bib/sdu-doi-general/main}

\end{document}


\graphicspath{{image/sdu-doi-general/ru/}}

\title{Операторный подход к основному кинетическому уравнению для одношаговых процессов}

\author{М. Гнатич}
\email{hnatic@saske.sk}
\affiliation{Лаборатория теоретической физики,\\
  Объединённый институт ядерных исследований,\\
  ул. Жолио-Кюри 6, Дубна, Московская область, Россия, 141980}

\author{Е. Г. Еферина}
\email{eg.eferina@gmail.com}
\affiliation{Кафедра прикладной информатики и теории вероятностей,\\
  Российский университет дружбы народов,\\
  ул. Миклухо-Маклая, д.6, Москва, Россия, 117198}

\author{А. В. Королькова}
\email{avkorolkova@gmail.com} 
\affiliation{Кафедра прикладной информатики и теории вероятностей,\\
  Российский университет дружбы народов,\\
  ул. Миклухо-Маклая, д.6, Москва, Россия, 117198}

\author{Д. С. Кулябов}
\email{yamadharma@gmail.com}
\affiliation{Кафедра прикладной информатики и теории вероятностей,\\
  Российский университет дружбы народов,\\
  ул. Миклухо-Маклая, д.6, Москва, Россия, 117198}
\affiliation{Лаборатория информационных технологий,\\
Объединённый институт ядерных исследований,\\
ул. Жолио-Кюри 6, Дубна, Московская область, Россия, 141980}

\author{Л. А. Севастьянов}
\email{leonid.sevast@gmail.com}
\affiliation{Кафедра прикладной информатики и теории вероятностей,\\
  Российский университет дружбы народов,\\
  ул. Миклухо-Маклая, д.6, Москва, Россия, 117198}
\affiliation{Лаборатория теоретической физики,\\
  Объединённый институт ядерных исследований,\\
  ул. Жолио-Кюри 6, Дубна, Московская область, Россия, 141980}

\begin{abstract}

  \begin{description}

  \item[Предпосылки]
Взгляд на вероятность как имманентное свойство природы приводит
исследователей к необходимости переходить от детерминистического
описания явлений к стохастическому.  На основе идеологии Н.~Г. ван
Кампена и К.~В. Гардинера была сформулирована методика стохастизации
одношаговых процессов. Она позволяет по виду уравнений кинетики
(уравнений взаимодействия) и предположению о характере процесса (это
может быть не обязательно процесс рождения"=гибели) записать основное
кинетическое уравнение.  Кинетика взаимодействия привлекает в
последнее время пристальное внимание к себе, поскольку часто возникает
в физических, химических, технических, биологических, экологических,
экономических, социологических системах.  Однако общие методы для
прямого исследования полученного уравнения отсутствуют. В методике
стохастизации одношаговых процессов используется разложение уравнения
в формальный ряд Тейлора (так называемое разложение
Крамерса--Мойала). Также возможно разложение по обратному объёму
системы (разложение ван Кампена). Оставляя в разложении члены вплоть
до второго порядка малости можно получить уравнение Фоккера--Планка, а
соответственно и уравнение Ланжевена. При этом нужно ясно понимать,
что эти уравнения являются приближённой записью основного
кинетического уравнения.

\item[Цель]
Однако это не снимает необходимости в методике исследования самого
основного кинетического уравнения. Тем более, что степенной ряд,
получаемый при разложении основного кинетического уравнения может
расходиться (например, при наличии зависимости модели от от
пространственных переменных), что делает невозможным применение
классической теории возмущений.

\item[Методы]
Выглядит перспективным применение квантово-полевой теории возмущений
для статистических систем с использованием частичных сумм (так
называемый метод Дои). При этом члены ряда теории возмущений
рассматриваются в духе фейнмановских интегралов по траекториям, где
роль пропагаторов играют функции Грина возмущённой части оператора
Лиувилля основного кинетического уравнения. Для большего удобства
выделения невозмущённой и возмущённой частей оператора Лиувилля, а
также получения явного вида функции Грина исходное уравнение
переписывается в представлении чисел заполнения (состояния Фока).

\item[Результаты]
Данная работа представляет собой методический материал, описывающий
принципы решения основного кинетического уравнения методом
квантово"=полевой теории возмущений. Особенность изложения состоит в
том, что оно рассчитано на неспециалистов в квантовой теории поля.  В
качестве демонстрационного примера предлагается модель Ферхюльста по
причине её простоты и наглядности (первого порядка, не зависит от
пространственных переменных, однако содержит нелинейность).

\item[Выводы]
Показана полная эквивалентность комбинаторного и операторного методов
получения и исследования основного кинетического уравнения для
одношаговых процессов.

  \end{description}

\end{abstract}

   \keywords{символьные методы в биологии: стохастические дифференциальные уравнения; основное кинетическое уравнения; 
     уравнение Фоккера--Планка; популяционные модели}

\maketitle

\section{Введение}

Для построения стохастических моделей одношаговых
процессов~\cite{ef-kor-gev-kul-sev:vestnik-miph:2014-3}
(процессов рождения гибели) была отработана комбинаторная методика на основе
идеологии Н.~Г. ван Кампена~\cite{van-kampen:stochastic::ru} и
К.~В. Гардинера~\cite{gardiner:stochastic::ru}. 
В рамках данной методики из схем взаимодействия строится основное
кинетическое уравнение (конкретно в нашей методике исследуются
одношаговые процессы). Само основное кинетическое уравнение не
исследуется, а, вместо этого, преобразуется к уравнению
Фоккера--Планка путём разложения в формальный ряд (разложение
Крамерса--Мойала)~\cite{gardiner:stochastic::ru}. Однако, возникает
необходимость обоснования возможности применения данного разложения
для каждого типа процессов. 

Таким образом возникает необходимость как исследования самого
основного кинетического уравнения, так и обоснование его
разложения. Представляется, что наилучшем образом этим требованиям
отвечает квантовая теория возмущений. 

Обычно, для квантовой теории возмущений используют два формализма:
формализм интегралов по траекториям и формализм вторичного
квантования (канонический формализм). По большому счёту, применение
одного из них есть дело вкуса. В целом ряде
работ~\cite{Doi:1976:second_quantization, Doi:1976:stochastic_theory,
  Hnatich:2000:velocity-fluctuation, hnatich:2011:random_sources::ru,
  hnatich:2011:anomalous_kinetics::ru,
  Hnatich:2013:field-theoretic_technique} было исследовано возможность
применения формализма вторичного квантования для статистических
задач. Однако стиль изложения в этих статьях направлен на физиков
теоретиков, что сильно ограничивает аудиторию, способную
воспользоваться данными наработками.

Авторы ставили себе задачу изложить методику применения операторного
метода хоть и формально, но доступно для первоначального изучения
неспециалисту в соответствующих областях теоретической физики.

Структура статьи следующая. В разделе \ref{sec:notation} введены
основные обозначения и соглашения. В разделе~\ref{sec:one-step} даётся
краткое введение в метод стохастизации одношаговых процессов.
В разделе~\ref{sec:fock} описывается алгоритм записи одношаговых
процессов в представлении чисел заполнения. А также записывается
основное кинетическое уравнения в форме операторного уравнения Лиувилля.
В разделе~\ref{sec:model} описывается иллюстративная модель, на базе
которой демонстрируются как комбинаторный, так и операторный
подходы. Показывается их эквивалентность.

\section{Обозначения и соглашения}
\label{sec:notation}

\begin{enumerate}

\item В работе используется нотация абстрактных
  индексов~\cite{penrose-rindler:spinors::ru}. В данной нотации тензор как
  целостный объект обозначается просто индексом (например, $x^{i}$),
  компоненты обозначаются подчёркнутым индексом (например,
  $x^{\crd{i}}$).

\item Будем придерживаться следующих соглашений. Латинские индексы из
  середины алфавита ($i$, $j$, $k$) будут относиться к пространству
  векторов состояний системы. Латинские индексы из начала алфавита
  ($a$) будут относиться к пространству винеровского
  процесса. Греческие индексы ($\alpha$) будут задавать количество
  разных взаимодействий в кинетических уравнениях.

\end{enumerate}

\section{Стохастизация одношаговых процессов}
\label{sec:one-step}

Под одношаговыми процессами (также известные как процессы
рождения--гибели) понимаются марковские процессы с непрерывным
временем, принимающие значения в области целых чисел, матрица перехода
которых допускает только переходы между соседними состояниями.

\subsection{Схемы взаимодействия}

Состояние системы будем описывать вектором состояния
$\varphi^{i} \in \setR^n$, где $n$~--- размерность
системы\footnote{Для краткости мы обозначаем модуль над полем $\setR$
  просто как $\setR$.}. Оператор
$I^{i}_{j} \in \setN^{n}_{0} \times \setN^{n}_{0}$ задаёт состояние
системы до взаимодействия, оператор
$F^{i}_{j} \in \setN^{n}_{0} \times \setN^{n}_{0}$~---
после\footnote{Соответственно, компонентные индексы размерности
  системы пробегают значения $\crd{i},\crd{j} = \overline{1,n}$}. В результате
взаимодействия происходит переход системы в другое состояние.

В системе может происходить $s$ видов различных
взаимодействий. Поэтому вместо операторов $I^{i}_{j}$ и $F^{i}_{j}$
будем рассматривать операторы
$I^{i \alpha}_{j} \in \setN^{n}_{0} \times \setN^{n}_{0} \times
\setN^{s}_{+}$
и
$F^{i \alpha}_{j} \in \setN^{n}_{0} \times \setN^{n}_{0} \times
\setN^{s}_{+}$\footnote{Соответственно,
  компонентные индексы количества взаимодействий пробегают значения
  $\crd{\alpha} = \overline{1,s}$}.

Взаимодействие элементов системы будем описывать с помощью схем
взаимодействия, подобным схемам химической кинетики:
\begin{equation}
  \label{eq:chemkin}
  I^{i \alpha}_{j} \varphi^j
  \overset{\tensor*[^{+}]{k}{_{\alpha}}}{\underset{\tensor*[^{-}]{k}{_{\alpha}}}{\rightleftharpoons}}
  F^{i \alpha}_{j} \varphi^{j},
\end{equation}
здесь греческие индексы задают количество взаимодействий, а латинские~---
размерность системы. Коэффициенты $\tensor*[^{+}]{k}{_{\alpha}}$ и
$\tensor*[^{-}]{k}{_{\alpha}}$ имеют смысл интенсивности (скорости) взаимодействия.

Изменение состояния будет задаваться оператором
\begin{equation}
  \label{eq:r_i}
  r_j^{i \alpha} = F_j^{i \alpha} -I_j^{i \alpha}.
\end{equation}

Таким образом, один шаг взаимодействия $\crd{\alpha}$ в прямом и
обратном направлениях можно записать соответственно как
\begin{equation}
  \begin{gathered}
    \varphi^{i}  \rightarrow \varphi^i + r^{i \crd{\alpha}}_{j} \varphi^{j},\\
    \varphi^{i} \rightarrow \varphi^{i} - r^{i \crd{\alpha}}_{j} \varphi^{j}.
  \end{gathered}
\end{equation}

Мы также можем записывать~\eqref{eq:chemkin} не в виде векторных уравнений, 
а в виде более традиционных сумм:
\begin{equation}
  \label{eq:chemkin2}
  I^{i \alpha}_{j} \varphi^j \delta_i
  \overset{\tensor*[^{+}]{k}{_{\alpha}}}{\underset{\tensor*[^{-}]{k}{_{\alpha}}}{\rightleftharpoons}}
  F^{i \alpha}_{j} \varphi^{j} \delta_i,
\end{equation}
где $\delta_{\crd{i}} = (1,\ldots,1)$.

Также мы будем использовать следующие обозначения:
\begin{equation}
  \label{eq:n^i-notion}
  I^{i \alpha} := I^{i \alpha}_{j} \delta^{j}, 
  \quad F^{i \alpha} := F^{i \alpha}_{j} \delta^{j}, \quad
  r^{i \alpha} := r^{i \alpha}_{j} \delta^{j}.
\end{equation}

\subsection{Основное кинетическое уравнение}

Для описания системы используется основное
кинетическое уравнение (\emph{master equation}), определяющее
вероятность перехода для марковского
процесса~\cite{van-kampen:stochastic::ru, gardiner:stochastic::ru}:
\begin{multline}
  \label{eq:master:trans}
  \pdv{p(\varphi_{2},t_{2}|\varphi_{1},t_{1})}{t} = \int \biggl[
  w(\varphi_{2}|\psi,t_{2}) p(\psi,t_{2}|\varphi_{1},t_{1}) 
  - {} \\ {} -
  w(\psi|\varphi_{2},t_{2}) p(\varphi_{2},t_{2}|\varphi_{1},t_{1}) 
  \biggr] \dd{\psi},
\end{multline}
где $w(\varphi|\psi,t)$ есть вероятность перехода из состояния $\psi$
в состояние $\varphi$ за единичное время.

Зафиксировав начальные значения $\varphi_{1},t_{1}$, можно записать 
данное уравнение для подансамбля:
\begin{equation}
  \label{eq:master:subansemble}
  \pdv{p(\varphi,t)}{t} = \int
  \qty[
  w(\varphi|\psi,t) p(\psi,t) -
  w(\psi|\varphi,t) p(\varphi,t)
  ] \dd{\psi}.
\end{equation}

При дискретной области определения $\varphi$ можно
записать~\eqref{eq:master:subansemble} (пронумеровав состояния числами
$n$ и $m$):
\begin{equation} 
  \label{eq:mas_eq}
    \pdv{p_{n}(t)}{t} = \sum\limits_{m} 
    \qty[w_{nm} p_{m}(t) - w_{mn} p_{n}(t)],
\end{equation}
где $p_{n}$~--- вероятность нахождения
системы в состоянии $n$ в момент времени $t$,
$w_{nm}$~--- вероятность перехода системы из состояния $m$ в
состояние $n$ в единицу времени.

В системе, описываемой одношаговыми процессами, возможны два вида
перехода системы из одного состояния в другое, происходящие в
результате взаимодействия элементов в прямом направлении
($\varphi^{i} + r^{i \crd{\alpha}}_{j} \varphi^{j}$) с вероятностью
$\tensor*[^{+}]{s}{_{\crd{\alpha}}}(\varphi^{k})$ и в обратном
направлении ($\varphi^{i} - r^{i \crd{\alpha}}_{j} \varphi^{j}$) с
вероятностью
$\tensor*[^{-}]{s}{_{\crd{\alpha}}}(\varphi^{k})$~(рис.~\ref{fig:one-step_process}). А
матрица вероятностей переходов может быть записана в виде:
\begin{equation}
  w_{\crd{\alpha}}(\varphi^{i}| \psi^{i} ,t) = \tensor*[^{+}]{s}{_{\crd{\alpha}}}
  \delta_{\varphi^{i},\psi^{i}+1} + \tensor*[^{-}]{s}{_{\crd{\alpha}}} \delta_{\varphi^{i}, \psi^{i}-1},
\end{equation}
где $\delta_{i,j}$ --- символ Кронекера.

Таким образом, общий вид основного кинетического уравнения для вектора
состояний $\varphi^{i}$, изменяющегося шагами длины $r^{i \crd{\alpha}}_j
\varphi^j$, принимает вид:
\begin{multline} 
  \label{eq:one-step:mu}
  \pdv{p(\varphi^{i} ,t)}{t} =  \sum_{\crd{\alpha}=1}^{s} 
  \Biggl\{
  \biggl[ \tensor*[^{-}]{s}{_{\crd{\alpha}}} (\varphi^{i}+r^{i
    \crd{\alpha}},t) p(\varphi^{i}+r^{i \crd{\alpha}} ,t) 
  - {} \\ - {}
  \tensor*[^{+}]{s}{_{\crd{\alpha}}}(\varphi^{i}) p(\varphi^{i},t)
  \biggr] 
  + {} \\ + {}
  \biggl[ \tensor*[^{+}]{s}{_{\crd{\alpha}}} (\varphi^{i}-r^{i
    \crd{\alpha}}, t) 
  p(\varphi^{i} - r^{i \crd{\alpha}},t) -
  \tensor*[^{+}]{s}{_{\crd{\alpha}}} (\varphi^i) p(\varphi^{i},t)
  \biggr] \Biggr\}.
\end{multline}

\begin{figure*}
  \centering
  \includegraphics[width=0.8\textwidth]{markov}
  \caption{Одношаговый процесс}
  \label{fig:one-step_process}
\end{figure*}

Интенсивности перехода в единицу времени
$\tensor*[^{+}]{s}{_{\crd{\alpha}}}$ и $\tensor*[^{-}]{s}{_{\crd{\alpha}}}$ пропорциональны соответственно числу
способов выбора числа размещений из $\varphi^{\crd{i}}$ по $I^{\crd{i}
  \crd{\alpha}}$ (обозначается как $A_{\varphi^{\crd{i}}}^{I^{\crd{i} \crd{\alpha}}}$) и по
$F^{\crd{i} \crd{\alpha}}$ (обозначается как $A_{\varphi^{\crd{i}}}^{F^{\crd{i} \crd{\alpha}}}$) и определяются выражениями:
\begin{equation}
  \label{eq:s-pm}
\begin{gathered}
  \tensor*[^{+}]{s}{_{\crd{\alpha}}} =
  \tensor*[^{+}]{k}{_{\crd{\alpha}}} \prod_{\crd{i}=1}^{n} 
  A_{\varphi^{\crd{i}}}^{I^{\crd{i} \crd{\alpha}}} =
  \tensor*[^{+}]{k}{_{\crd{\alpha}}} \prod_{\crd{i}=1}^{n}
  \frac{\varphi^{\crd{i}}!}{(\varphi^{\crd{i}} - I^{\crd{i} \crd{\alpha}})!}, \\
  \tensor*[^{-}]{s}{_{\crd{\alpha}}} = 
  \tensor*[^{-}]{k}{_{\crd{\alpha}}} \prod_{\crd{i}=1}^{n} 
  A_{\varphi^{\crd{i}}}^{F^{\crd{i} \crd{\alpha}}} =
  \tensor*[^{-}]{k}{_{\crd{\alpha}}} \prod_{\crd{i}=1}^{n}
  \frac{\varphi^{\crd{i}}!}{(\varphi^{\crd{i}}-F^{\crd{i} \crd{\alpha}})!}.
\end{gathered}
\end{equation}

Заменяя в~\eqref{eq:s-pm} комбинации типа
$\varphi (\varphi-1) \cdots (\varphi - (n-1))$ на $(\varphi)^n$
получим для уравнения Фоккера--Планка\footnote{Такая замена
  соответствует разложению в ряд.}:
\begin{equation}
  \label{eq:s-pm:exp}
\begin{gathered}
  \tensor*[^{+}_{\text{fp}}]{s}{_{\crd{\alpha}}} =
  \tensor*[^{+}]{k}{_{\crd{\alpha}}} \prod_{\crd{i}=1}^{n}
  (\varphi^{\crd{i}})^{I^{\crd{i} \crd{\alpha}}}, \\
  \tensor*[^{-}_{\text{fp}}]{s}{_{\crd{\alpha}}} = 
  \tensor*[^{-}]{k}{_{\crd{\alpha}}} \prod_{\crd{i}=1}^{n}
  (\varphi^{\crd{i}})^{F^{\crd{i} \crd{\alpha}}}.
\end{gathered}
\end{equation}

\subsection{Уравнение Фоккера--Планка}

Уравнение Фоккера--Планка является частным случаем основного
кинетического уравнения и может рассматриваться как его приближённая
форма.

Будем использовать разложение Крамерса--Мойала~\cite{gardiner:stochastic::ru} (для простоты записано
в одномерном случае):
\begin{equation}
  \label{eq:kramers-moyal}
  \pdv{p(\varphi,t)}{t} = \sum \limits_{n=1}^\infty \frac{(-1)^n}{n!}
  \pdv[n]{}{\varphi} \Big[\xi^n(\varphi) p(\varphi,t)\Big],
\end{equation}
где
\begin{equation}
  \xi^n(\varphi) = \int\limits_{-\infty}^{\infty} (\psi - \varphi)^{n}
  w(\psi|\varphi) \dd{\psi}.
\end{equation}

Отбрасывая члены выше второго порядка, получаем уравнение
Фоккера--Планка:
\begin{equation}
  \label{eq:fp_1D}
  \pdv{p(\varphi, t)}{t} = 
  - \pdv{\varphi} \left[ A(\varphi) p(\varphi, t) \right] +
  \pdv[2]{}{\varphi} 
  \left[ B(\varphi) p(\varphi, t) \right],
\end{equation}
или в многомерном случае:
\begin{multline}
  \label{eq:fp_nD}
  \pdv{p(\varphi^k, t)}{t} = 
  - \pdv{}{\varphi^{i}} \left[ A^{i}(\varphi^{k}) p(\varphi^{k}, t)
  \right] 
  + {} \\ {} +
  \frac{1}{2} \pdv{}{\varphi^{i}}{\varphi^{j}} 
  \left[ B^{i j} (\varphi^k) p(\varphi^{k}, t)  \right],
\end{multline}
где
\begin{equation} 
  \label{eq:fp_coeff}
  \begin{gathered}
    A^{i} := A^{i}(\varphi^{k}) = r^{i \crd{\alpha}} \left[
      \tensor*[^+_{\text{fp}}]{s}{_{\crd{\alpha}}} -
      \tensor*[^-_{\text{fp}}]{s}{_{\crd{\alpha}}} \right], \\
    B^{i j} := B^{i j}(\varphi^{k}) = r^{i \crd{\alpha}} r^{j
      \crd{\alpha}} \left[ \tensor*[^+_{\text{fp}}]{s}{_{\crd{\alpha}}} -
      \tensor*[^-_{\text{fp}}]{s}{_{\crd{\alpha}}} \right].
  \end{gathered}
\end{equation}

Как видно из~\eqref{eq:fp_coeff}, коэффициенты уравнения
Фоккера--Планка можно получить сразу из \eqref{eq:r_i} и
\eqref{eq:s-pm}, то есть в данном случае записывать основное
кинетическое уравнение нет необходимости.

\subsection{Уравнение Ланжевена}

Уравнению Фоккера--Планка соответствует уравнение Ланжевена:
\begin{equation}
  \label{eq:langevin}
  \dd \varphi^{i} = a^{i} \dd{t} + b^i_{a} \dd{W^{a}},
\end{equation}
где $a^{i} := a^{i} (\varphi^k)$, $b^{i}_{a} := b^{i}_{a}
(\varphi^k)$, 
$\varphi^i \in \setR^n $ --- вектор состояния системы, $W^{a} \in \mathbb{R}^m$
--- $m$-мерный винеровский процесс\footnote{Винеровский процесс реализуется
как $\dd{W} = \varepsilon \sqrt{\dd{t}}$, где $\varepsilon \sim
N(0,1)$~--- нормальное распределение со средним $0$ и дисперсией
$1$.}. Здесь латинскими индексами из середины алфавита обозначаются величины,
относящиеся к векторам состояний (размерность пространства $n$), а
латинскими индексами из начала алфавита обозначаются величины,
относящиеся к вектору винеровского процесса (размерность пространства
$m \leqslant n$).

При этом связь между уравнениями \eqref{eq:fp_nD} и \eqref{eq:langevin}
выражается следующими соотношениями:
\begin{equation}
  \label{eq:k-langevin}
  A^{i} = a^{i}, \qquad B^{i j} = b^{i}_{a} b^{j a}.
\end{equation}

\section{Представление чисел заполнения}
\label{sec:fock}

Представление чисел заполнение является основным языком при описании
физики многих тел.  Главными элементами этого языка являются волновые
функции системы, содержащие информацию о том, сколько частиц находится
в каждом одночастичном состоянии.  Для изменения состояния системы
используют операторы рождения и уничтожения. 
Преимущества данного формализма следующие:
\begin{itemize}
\item возможно рассматривать системы с переменным числом частиц
  (нестационарные системы);
\item статистика системы (Ферми--Дирака или Бозе--Эйнштейна)
  автоматически учитывается в правилах коммутации для операторов
  рождения--уничтожения;
\item данный формализм является вторым основным формализмом (наряду с
  интегралами по траекториям) для описания квантовой теории возмущений.
\end{itemize}

Методика применения формализма вторичного квантования для неквантовых
систем (статистических, детерминированных) была рассмотрена в целом
ряде статей~\cite{Doi:1976:second_quantization,
  Doi:1976:stochastic_theory, zeldovich:1978:kinetics::ru,
  grassberger:1980:fock-space,
  peliti:1985:path_integral_approach}.

Для записи представления чисел заполнения обычно используют нотацию
Дирака.

\subsection{Нотация Дирака}

В нотации, предложенной 
П. А. М. Дираком~\cite{dirac:1939:braket}\footnote{Нотация Дирака
  базируется на нотации, предложенной Г. Грассманом в 1862
  году~\cite[с.~134]{cajori:1929:history}.}, вектор
$\varphi^{i}$ задаётся в виде $\ket{i}$, а ковектор $\varphi_{i}$ в
виде $\bra{i}$. Операция сопряжения служит для подымания и опускания
индексов\footnote{В данном случае мы используем эрмитово сопряжение
  ${\bullet}^{\dagger}$. Знак комплексного сопряжения ${\bullet}^{*}$
  в данной записи, по большому счёту, излишен.}:
\begin{equation}
  \label{eq:dirac:index_conj}
  \varphi^{*}_{i} := \varphi_{i} = (\varphi^{i})^{\dagger} \equiv \bra{i} = \ket{i}^{\dagger}.
\end{equation}

Скалярное произведение имеет следующий вид:
\begin{equation}
  \label{eq:dirac:inner}
  \varphi_{i} \varphi^{i} \equiv \bra{i}\ket{i}.
\end{equation}

Тензорное произведение имеет вид:
\begin{equation}
  \label{eq:dirac:outer}
  \varphi_{j} \varphi^{i} \equiv \ket{i}\bra{j}.
\end{equation}

Кроме того, возможна нотация и следующего вида:
\begin{equation}
  \label{eq:dirac:wavefunc}
  \ket{\varphi} := \varphi^{i}, \qquad 
  \bra{i}\ket{\varphi} := \varphi^{i} \delta^{\crd{i}}_{i} = \varphi^{\crd{i}}.
\end{equation}

\subsection{Операторы рождения--уничтожения}

Переход к пространству чисел заполнения не является унитарным
преобразованием. Однако мы можем построить алгоритм перехода (впрочем,
специфичный для каждой задачи).

Запишем основное кинетическое уравнение~\eqref{eq:mas_eq} в
представлении чисел заполнения.  Таким образом, мы будем рассматривать
систему, не зависящую от пространственных переменных. Для простоты
рассмотрим одномерный вариант.

Обозначим в~\eqref{eq:mas_eq} через $\varphi_{n}$ вероятность
обнаружить в системе $n$ частиц:
\begin{equation}
  \label{eq:phi_n}
  \varphi_{n} := p_{n} (\varphi, t).
\end{equation}

Состояния $\varphi$ образуют векторное пространство $\mathcal{H}$.

Введём скалярное произведение:
\begin{equation}
  \label{eq:scalar_prod_ex}
  \bra{\varphi}\ket{\psi} = \sum_n n! p_{n}^{*} (\varphi) p^{n}(\psi)
  = \sum_n n! \varphi_{n}^{*} (\varphi) \varphi^{n}(\psi)
\end{equation}
и базисные векторы $\ket{n}$.

Из $p_{n}(m) = \delta_{n}^{m}$ и~\eqref{eq:scalar_prod_ex} следует:
\begin{equation}
  \label{eq:<n|m>}
\bra{n}\ket{m} = n!\delta_{n}^{m}.
\end{equation}

Запишем вектор состояния:
\begin{equation}
  \label{eq:phi_vector}
  \ket{\varphi} = \sum_{n} p_{n} (\varphi) \ket{n} = \sum_{n}
  \varphi_{n} \ket{n} =: \varphi_{n} \ket{n}.
\end{equation}

С учётом~\eqref{eq:<n|m>} имеем:
\begin{equation}
  \label{eq:phi_n-scalar}
  \varphi_{n} = \frac{1}{n!} \bra{n}\ket{\varphi}.
\end{equation}

Введём операторы рождения и уничтожения:
\begin{equation}
  \label{eq:creat+annihil}
  \begin{gathered}
    \pi \ket{n} = \ket{n+1}, \\
    a \ket{n} = n \ket{n-1}.
  \end{gathered}
\end{equation}
Их коммутационное соотношение имеет вид\footnote{Действительно, $a\pi\ket{n} -
  \pi a \ket{n} = (n+1)\ket{n} - n \ket{n} = \ket{n}$.}:
\begin{equation}
  \label{eq:commutator}
  [a,\pi] = 1.
\end{equation}

Из~\eqref{eq:commutator} видно, что при выборе скалярного произведения
в виде~\eqref{eq:scalar_prod_ex} система описывается статистикой
Бозе--Эйнштейна. 

Из соотношения~\eqref{eq:<n|m>} получим:
\begin{equation}
  \label{eq:scalar-complex-conj}
  \mel{m}{a^{\dagger}}{n} = \mel{m}{\pi}{n},
\end{equation}
таким образом, для скалярного произведения~\eqref{eq:scalar_prod_ex}
имеем
\begin{equation}
  \label{eq:a+=pi}
  a^{\dagger} = \pi.
\end{equation}

\subsection{Оператор Лиувилля}

Запишем уравнение Лиувилля:
\begin{equation}
  \label{eq:liuville}
  \pdv{}{t} \ket{\varphi(t)} =  L \ket{\varphi(t)}.
\end{equation}
Оператор Лиувилля $L$ удовлетворяет соотношению:
\begin{equation}
  \label{eq:liuville=0}
  \bra{0} L = 0.
\end{equation}

Из~\eqref{eq:mas_eq}, \eqref{eq:phi_vector}, \eqref{eq:phi_n-scalar}
и~\eqref{eq:liuville} получаем:
\begin{multline}
  \label{eq:master2liuville}
   \pdv{p_{n}}{t} = \frac{1}{n!} \bra{n}\pdv{}{t}\ket{\varphi} =
   \frac{1}{n!} \bra{n}L\ket{\varphi} 
   = {} \\ {} =
   \sum\limits_{m} 
    \qty[w_{nm} p_{m} - w_{mn} p_{n}].
\end{multline}
То есть уравнение Лиувилля~\eqref{eq:liuville} в форме одного
уравнения записывает набор основных кинетических
уравнений~\eqref{eq:mas_eq} для разных значений $n$.

Схеме~\eqref{eq:chemkin} соответствует оператор Лиувилля:
\begin{multline}
  \label{eq:liuville:chemkin}
  L = \sum_{\crd{\alpha},\crd{i}} \biggl[
  \tensor*[^{+}]{k}{_{\crd{\alpha}}}
  \qty(
  (\pi_{\crd{i}})^{F^{\crd{i}\crd{\alpha}}} -
  (\pi_{\crd{i}})^{I^{\crd{i}\crd{\alpha}}} 
  ) (a_{\crd{i}})^{I^{\crd{i}\crd{\alpha}}}
  + {} \\ {} +
  \tensor*[^{-}]{k}{_{\crd{\alpha}}}
  \qty(
  (\pi_{\crd{i}})^{I^{\crd{i}\crd{\alpha}}} -
  (\pi_{\crd{i}})^{F^{\crd{i}\crd{\alpha}}} 
  ) (a_{\crd{i}})^{F^{\crd{i}\crd{\alpha}}}
  \biggr].
\end{multline}

\section{Элементы теории возмущений}

\section{Модель Ферхюльста}
\label{sec:model}

В качестве демонстрации метода рассмотрим модель
Ферхюльста~\cite{verhulst:1838, Feller:1939:acta_biotheoretica,
  feller:1949:theory_stochastic_processes}, описывающую ограниченный
рост\footnote{Привлекательность этой модели в том, что она одномерна и
  нелинейна.}. Изначально эта модель описывается следующим
дифференциальным уравнением:
\begin{equation}
    \dv{\varphi}{t}= \lambda \varphi - \beta \varphi - \gamma \varphi^{2},
\end{equation}
здесь $\lambda$~--- коэффициент интенсивности размножения, $\beta$~---
коэффициент интенсивности вымирания, $\gamma$~--- коэффициент
интенсивности уменьшения популяции (обычно рассматривается
соперничество особей)\footnote{Здесь мы оставляем те же обозначения,
  что и в исходной модели~\cite{verhulst:1838}.}.

Построим стохастический вариант данной модели. Запишем схемы
взаимодействия:
\begin{equation}
  \label{eq:verhulst:chemkin}
    \begin{gathered}
        \varphi \overset{\lambda}{\underset{\gamma}{\rightleftharpoons}} 2\varphi ,\\
        \varphi \xrightarrow{\beta} 0.
    \end{gathered}
    \qquad
    \begin{gathered}
      I^{\crd{i}\crd{\alpha}}=
      \begin{pmatrix}
        1 & 1 
      \end{pmatrix}, \\
      F^{\crd{i}\crd{\alpha}}=
      \begin{pmatrix}
        2 & 0 
      \end{pmatrix}.
    \end{gathered}
    \qquad
    \begin{gathered}
      r^{\crd{i}\crd{\alpha}} =
      \begin{pmatrix}
        1 & -1
      \end{pmatrix}.
    \end{gathered}
  \end{equation}

Первое соотношение означает, что индивидуум, который съедает единицу
пищи, немедленно репродуцируется, в обратную сторону~--
соперничество между индивидуумами. Второе~---  смерть индивидуума.

\subsection{Метод стохастизации одношаговых процессов}

Согласно~\eqref{eq:s-pm:exp} определим интенсивности перехода:
\begin{equation}
  \begin{gathered}
    \tensor*[^{+}]{s}{_{1}} = \lambda \varphi, \\
    \tensor*[^{-}]{s}{_{1}} = \gamma \varphi (\varphi - 1), \\
    \tensor*[^{+}]{s}{_{2}} = \beta \varphi.
  \end{gathered}
  \qquad
  \begin{gathered}
    \tensor*[^{+}_{\text{fp}}]{s}{_{1}} = \lambda \varphi, \\
    \tensor*[^{-}_{\text{fp}}]{s}{_{1}} = \gamma \varphi^{2}, \\
    \tensor*[^{+}_{\text{fp}}]{s}{_{2}} = \beta \varphi.
  \end{gathered}
\end{equation}

Тогда (на основании~\eqref{eq:one-step:mu}) основное кинетическое
уравнение примет следующий вид:
\begin{multline}
  \label{eq:master}
  \pdv{p (\varphi,t)}{t} =  
  - \qty[\lambda \varphi + \beta \varphi +
  \gamma \varphi (\varphi -1)] 
  p(\varphi,t) 
  + {} \\ {} +
  \qty[\beta (\varphi + 1)
  +  \gamma (\varphi + 1) \varphi] 
  p(\varphi + 1,t) + \lambda (\varphi - 1) p (\varphi - 1,t).
\end{multline}

Или же, записывая для конкретных значений $\varphi$ (как
в~\eqref{eq:mas_eq}):
\begin{multline}
  \label{eq:verhulst:master:n}
  \pdv{p_{n}(t)}{t} := \pdv{p (\varphi,t)}{t}\eval_{\varphi=n} 
  = {} \\ {} =
  - \qty[\lambda n + \beta n +
  \gamma n(n-1)] 
  p_{n}(t) 
  + {} \\ {} +
  \qty[\beta (n + 1) + \gamma (n + 1)n] 
  p_{n+1}(t) + 
  \lambda (n - 1) p_{n-1} (t).
\end{multline}

Пользуясь формулой~\eqref{eq:fp_1D}, получим уравнение
Фоккера--Планка:
\begin{equation}
  \pdv{p(\varphi,t)}{t}=
  -\pdv{}{\varphi} \left(A p(\varphi,t)\right)+\frac{1}{2}
  \pdv[2]{}{\varphi} \left(B p(\varphi,t)\right),
\end{equation}

Коэффициенты $A$ и $B$ соответственно равны (см.~\eqref{eq:fp_coeff}):

\begin{equation}
  \begin{gathered}
    A = \lambda \varphi -\beta\varphi - \gamma \varphi^2, \\
   B = \lambda \varphi + \beta\varphi - \gamma \varphi^2.
  \end{gathered}
\end{equation}

Из уравнения Фоккера--Планка получим эквивалентное ему
СДУ в форме Ланжевена:
\begin{equation}
  \dd{\varphi(t)} = (\lambda \varphi -\beta\varphi - \gamma
  \varphi^2)\dd{t} +
 \sqrt {(\lambda \varphi + \beta\varphi - \gamma \varphi^2)} \dd{W(t)}.
\end{equation}

\subsection{Представление чисел заполнения}

На основании~\eqref{eq:verhulst:chemkin} и~\eqref{eq:liuville:chemkin}
получаем оператор Лиувилля:
\begin{multline}
  \label{eq:verhulst:liuville}
  L = \lambda (\pi^2 -\pi) a + \gamma (\pi - \pi^2) a^2 + 
  \beta (1 - \pi) a 
  = {} \\ {} =
  \lambda \qty((a^{\dagger})^2 - a^{\dagger}) a + 
  \gamma \qty(a^{\dagger} - (a^{\dagger})^2) a^2 + 
  \beta \qty(1 - a^{\dagger}) a 
  = {} \\ {} =
  \lambda \qty(a^{\dagger} - 1)a^{\dagger} a +
  \beta \qty(1 - a^{\dagger}) a + 
  \gamma \qty(1 - a^{\dagger})a^{\dagger} a^2.
\end{multline}

Запишем основное кинетическое уравнение через оператор Лиувилля (из
формулы~\eqref{eq:master2liuville}) и учтём~\eqref{eq:<n|m>}, \eqref{eq:creat+annihil},
\eqref{eq:phi_n} и~\eqref{eq:phi_n-scalar}:
\begin{multline}
  \label{eq:verhulst:master2liuville}
  \pdv{p_{n}(t)}{t} = \frac{1}{n!} \bra{n}L\ket{\varphi} 
  =  {} \\ {} =
  \frac{1}{n!} \bra{n}
  - \qty[ 
  \lambda a^{\dagger} a +
  \beta a^{\dagger} a +
  \gamma a^{\dagger}a^{\dagger} a a
  ]
  + {} \\ {} +
  \qty[
  \beta a + 
  \gamma a^{\dagger} a a
  ]
  +
  \lambda a^{\dagger}a^{\dagger} a 
  \ket{\varphi}
  = {} \\ {} =
  - \qty[\lambda n + \beta n +
  \gamma n(n-1)] 
  \bra{n}\ket{\varphi}
  + {} \\ {} +
  \qty[\beta (n + 1) + \gamma (n + 1)n] 
  \bra{n+1}\ket{\varphi} + 
  \lambda (n - 1) \bra{n-1}\ket{\varphi}
  = {} \\ {} =
  - \qty[\lambda n + \beta n +
  \gamma n(n-1)] 
  p_{n}(t) 
  + {} \\ {} +
  \qty[\beta (n + 1) + \gamma (n + 1)n] 
  p_{n+1}(t) + 
  \lambda (n - 1) p_{n-1} (t).
\end{multline}

Результат~\eqref{eq:verhulst:master2liuville} полностью совпадает с
формулой~\eqref{eq:master}, полученной комбинаторным методом.

\section{Заключение}
\label{sec:conclusion}

Таким образом, в статье введён операторный метод для одношаговых
процессов. На всех этапах операторный метод сравнивается с
комбинаторным методом стохастизации одношаговых
процессов. Демонстрируется логика построения обоих методов. Сравнение
показывает их полную эквивалентность. 

Впрочем, на данном этапе обосновать предпочтение одного из методов
представляется трудной задачей. Хотя нельзя не отметить, что
операторный формализм позволяет более привычным образом использовать
наработки, сделанные в рамках квантовой теории поля.

\def\acknowledgmentsname{Благодарности}

\begin{acknowledgments}

Работа частично поддержана грантами РФФИ \textnumero~14-01-00628 и
15-07-08795. 

Опубликовано в:
Hnati{\v{c}}~M., Eferina~E.~G., Korolkova~A.~V. et~al. Operator
Approach to the Master Equation for the One-Step Process~//
EPJ Web of
Conferences.~--- 2016.~--- Vol. 108.~--- P.~02027.~---
\href{http://dx.doi.org/10.1051/epjconf/201610802027}{doi:10.1051/epjconf/201610802027}.

Исходные тексты:
\url{https://bitbucket.org/yamadharma/articles-2014-sdu-doi}

\end{acknowledgments}

  \ifthenelse{\boolean{xetex}}
  {\bibliographystyle{ugost2008l}}
  {\bibliographystyle{gost2008l}}

\bibliography{bib/sdu-doi-general/main}